\documentclass[allclo]{FBSart}
\usepackage{amsfonts}
\usepackage{amssymb}
\usepackage{epsfig}

\newcommand{\gk}{\kappa}

\newcommand{\gl}{\lambda}

\newcommand{\gL}{\Lambda}

\title{On the stability of three-body bound states on the light front} 

\author{M. Beyer$^a$\thanks{{\it E-mail address:}
    michael.beyer@physik.uni-rostock.de}, S.
  Mattiello$^a$\thanks{{\it E-mail address:}
    stefano.mattiello@physik.uni-rostock.de}, T.  Frederico$^b$, H. J.
  Weber$^c$}

\institute{$^a$Fachbereich Physik, Universit\"at Rostock, D-18051
  Rostock, Germany\\$^b$Dep. de F\'\i sica, Instituto Tecnol\'ogico de
  Aeron\'autica, Centro T\'ecnico Aeroespacial,
  12.228-900 S\~ao Jos\'e dos Campos, S\~ao Paulo, Brazil\\
  $^c$Dept. of Physics, University of Virginia, Charlottesville, VA
  22904, U.S.A.}

\runningauthor{M. Beyer}
\runningtitle{Stability of three-body bound states}
\begin{document}
 \maketitle
\begin{abstract}
  We investigate the stability of the relativistic three-boson system
  with a zero range force in the light front form. In particular we
  study the dependence of the system on an invariant cut-off. We
  discuss the conditions for the relativistic Thomas collapse.
  Finally, we fix the parameters of the model introducing a scale.\\
\end{abstract}
\section{Introduction}

Recently, there has been a renewed interest in zero range
interactions~\cite{Fedorov:2001wj,Fedorov:2001vw,Carbonell:2002qs}.
Zero range interactions provide a simple, but important limiting case
for short range forces.  As such, they are useful for many
applications in nuclear and atomic physics.  Furthermore, the zero
range interaction for nucleons has been revisited in the context of
effective chiral theories, see e.g.~\cite{Bedaque:2002mn}.  It is an
appealing approximation for few-body problems, as it simplifies the
corresponding equations~\cite{Noyes:1982ak}.  On the other hand, it is
well known that the non-relativistic three-body system based on zero
range forces experiences the Thomas collapse~\cite{Thomas1935}. To
prevent the Thomas collapse several regularization schemes have been
suggested~\cite{Fedorov:2001wj,Fedorov:2001vw,Adhikari:1994fc,Frederico1999}.

The Thomas collapse occurs when the binding energy of the system is
unbounded from below. However, as the binding energy becomes larger,
it eventually compares with and exceeds the size of the constituent
masses and hence, a nonrelativistic treatment of the problem is not
sufficient. The relativistic three-particle problem with zero range
interactions has been tackled using minimal relativistic
three-particle equations~\cite{Lindesay:1986kg} and relativistic light
front equations~\cite{Frederico:1992uw,deAraujo:1995mh}. The latter
has been revisited by Carbonell and Karmanov in a covariant light
front approach. Given a constituent mass $m$ of the particles they
succeed to solve the three-body equation without introducing a
regularization scheme~\cite{Carbonell:2002qs}. Previously, a
regularization procedure has been introduced
~\cite{Lindesay:1986kg,Frederico:1992uw} that is equivalent to a
smearing of the zero range interaction. Moreover, the particular
cut-off chosen in these approaches prevents the relativistic analog of
the Thomas collapse from occuring, namely the zero mass limit for the
three-body bound state. In this paper, we study the dependence on an
invariant cut-off $\Lambda$. The numerical limit $\Lambda\rightarrow
\infty$ coincides with the results of Carbonell and Karmanov, where a
comparison can be made. The relativistic analog of the Thomas collapse
appears for all $\Lambda$. We find that for weakly bound states the
dependence on the cut-off is mild.

\section{Theory}
We briefly review the basic ingredients of the three-body equations on
the light front given earlier~\cite{Frederico:1992uw,deAraujo:1995mh}.
\begin{figure}[t]
\begin{center}
\epsfig{figure=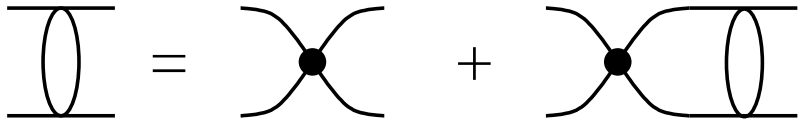,width=0.8\textwidth}
\end{center}
\caption{\label{fig:T2}
  Equation for the two-body $t$-matrix with zero
  range interaction.}
\end{figure}
For a zero range interaction in the particle-particle channel the
equation represented by Fig.~\ref{fig:T2} can be summed and leads to a
solution for the two-particle propagator $t(M_2)$, i.e.
\begin{equation}
t(M_2)=\left(i\gl^{-1} - B(M_2)\right)^{-1}.
\label{eqn:tau}
\end{equation}
The expression for $B(M_2)$ corresponds to a loop diagram. In the rest
system of the two-body system $P^\mu=(M_2,0,0,0)$ it is given by
\begin{equation}
B(M_2)=-\frac{i}{2(2\pi)^3} \int \frac{dx d^2k_\perp}{x(1-x)}
\frac{1}{M_2^2-M_{20}^2},
\label{eqn:B}
\end{equation}
where
\begin{eqnarray}
M_{20}^2&=&(\vec k_\perp^{~2}+m^2)/x(1-x).
\end{eqnarray}
The light front coordinates of particle 1 are $k^\pm,\vec k_\perp$,
where $k^\pm=k^0\pm k_z$, $x=k^+/P^+$, and $\vec k_\perp=(k_x,k_y)$.
The integral involving $B(M_2)$ has a logarithmic divergence that can
be absorbed in a redefinition of $\gl$. To do so one has to assume
that the two particle amplitude $t(M_2)$ has a pole for $M_2=M_{2B}$.
We may then write
\begin{equation}
i\gl^{-1}=B(M_{2B}).
\label{eqn:Bound}
\end{equation}
Hence, the physical information introduced in the renormalization of
the amplitude is the mass of the two bound particles, $M_{2B}$.  The
subtraction imposed by condition (\ref{eqn:Bound}) in the denominator
of eq.~(\ref{eqn:tau}) makes $t(M_2)$ finite. The resulting
expression for the two-body propagator is then given by
\begin{eqnarray}
t(M_2) &= &\Big(i
\left[\gk(M_{2B}) {\rm arctan}2\gk^{-1}(M_{2B})\right.
\nonumber\\
&&\left.-\gk(M_{2}) {\rm arctan}2\gk^{-1}(M_{2})\right]\Big)/(2\pi)^2,
\label{eqn:tau2}
\end{eqnarray}
where
\begin{equation}
\gk(M_2)= \sqrt{\frac{m^2}{M_2^2}-\frac{1}{4}}.
\end{equation}
The above eq. (\ref{eqn:Bound}) has been utilized
in~\cite{Frederico:1992uw} and subsequent
calculations~\cite{deAraujo:1995mh,Beyer:2001bc,Mattiello:2001vq,
Carbonell:2002qs}.
However, it is restricted to cases with a two-body bound state.  In
order to investigate a solution of the three-body bound state equation
also for cases where no two-body bound state exists, we now introduce
an invariant cut-off in the integral (\ref{eqn:B}), i.e.
\begin{equation}
M_{20}^2<\Lambda^2,
\end{equation}
which makes the integral finite. With this cut-off the integral reads
\begin{equation}
B_\Lambda(M_2)=-\frac{2\pi i}{2(2\pi)^3} 
\int\limits_{x_{\rm min}}^{x_{\rm max}}
 \frac{dx }{x(1-x)} \int\limits_0^{k_{\rm max}}k_\perp dk_\perp
\frac{1}{M_2^2-M_{20}^2},
\label{eqn:Breg}
\end{equation}
where
\begin{eqnarray}
x_{\rm min}&=&\frac{1}{2}\left(1-\sqrt{1-4m^2/\gL^2}\right)\\
x_{\rm max}&=&\frac{1}{2}\left(1+\sqrt{1-4m^2/\gL^2}\right)\\
k^2_{\rm max}&=&\gL^2x(1-x)-m^2
\end{eqnarray}
Consequently $t(M_2)\rightarrow t_\Lambda(M_2)$ depends on $\Lambda$ as well.

\begin{figure}
\begin{center}
    \epsfig{figure=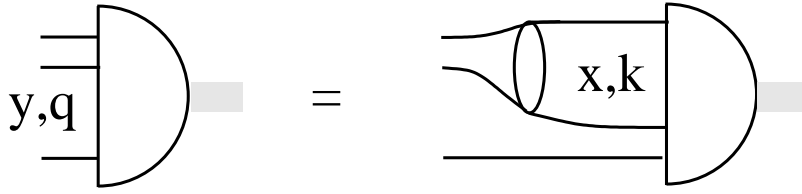,width=0.8\textwidth}
\end{center}
\caption{\label{fig:B3}
  Diagrammatic representation of the Faddeev equation for a zero range
  interaction, eq. (\ref{eqn:fad}). The two-body input is given in
  Fig.~\protect\ref{fig:T2} and eq.~(\protect\ref{eqn:tau}).}
\end{figure}
We now turn to the three-particle case. 
 The expression for the two-body mass embedded in the three-body system is
\begin{eqnarray}
M_2^2&=&(P_{3} - q)^2\nonumber\\
&=&(M_3-q^+)\left(M_3-\frac{\vec q_\perp^{~2}+m^2}{q^+}\right)
-\vec q_\perp^{~2},
\label{eqn:M2}
\end{eqnarray}
where $q$ denotes the momentum of the third particle.  The three-body
system is taken at rest, $P^\mu_3=(M_3,0,0,0)$.  In the three-body
rest frame we define $x=k^+/M_3$ and $y=q^+/M_3$. Again we formally
introduce an invariant regularization,
\begin{equation}
M^2_{30}<\Lambda^2.
\end{equation}
Hence the relativistic equation (see Fig. \ref{fig:B3}) on
the light front is given by
\begin{eqnarray}
\Gamma_\Lambda(y,\vec q_\perp) &= &\frac{i}{(2\pi)^3}\ t_\Lambda(M_2)
\int_{0}^{1-y} \frac{dx}{x(1-y-x)}\nonumber\\
&&\int d^2k_\perp
\frac{\theta(M^2_{30}-\Lambda^2)}
{M^2_3 -M_{03}^2}\;\Gamma_\Lambda(x,\vec k_\perp),
\label{eqn:fad}
\end{eqnarray}
where we have introduced the vertex function $\Gamma_\Lambda$.  The mass
of the virtual three-particle state (in the rest system) is
\begin{equation}
M_{30}^2=\frac{\vec k^{~2}_\perp+m^2}{x}
+\frac{\vec q^{~2}_\perp+m^2}{y}
+\frac{(\vec k+\vec q)^2_\perp+m^2}{1-x-y},
\end{equation}
which is the sum of the on-shell minus-components of the three
particles.  

Equation (\ref{eqn:fad}) coincides with the one given
previously \cite{Frederico:1992uw} except for the regularization
scheme, which we call scheme A in the following.  In
\cite{Frederico:1992uw} the integration limits have been introduced to
satisfy $M_2^2>0$ and hence
\begin{equation}
\int_{0}^{1-y} dx\int d^2k_\perp
\rightarrow \int_{M^2/M_3^2}^{1-y} dx
\int^{k_\perp^{\mathrm{max}}} d^2k_\perp
\end{equation}
with $k_\perp^{\mathrm{max}}=\sqrt{(1-x)(xM_3^2-m^2)}$. As noted
correctly by ~\cite{Carbonell:2002qs} in this case one no longer deals
with zero range forces. In combination with (\ref{eqn:Bound}) the
invariant cut-off chosen here allows us to perform the zero range
limit for $\Lambda\rightarrow\infty$. However, beyond that, the
cut-off allows us to treat three-body states without two-body bound states as
well, i.e. where (\ref{eqn:Bound}) does not hold.

\section{Results}

We present our results in two subsections. First we focus on the more
general aspects of the calculation without fixing any of the
parameters. Later on we choose the parameters to reproduce the
physical proton mass (as in Ref.~\cite{deAraujo:1995mh}) to discuss
more quantitative aspects of the approach.

\begin{figure}[ht]
\begin{center}
    \epsfig{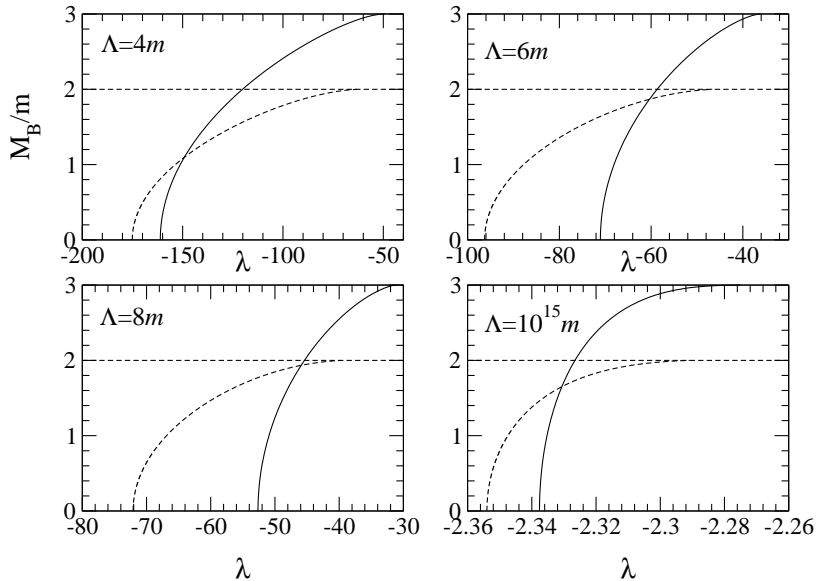}
\end{center}
\caption{\label{fig:m23L}
  Solution of the two- (dashed) and three-body (solid) bound state
  equations as a function of the strength $\lambda_\Lambda$ for
  different cut-off parameters $\Lambda=4m,6m,8m$ and
  $\Lambda=10^{15}m$ Figs. a,b,c, and d respectively.  Horizontal
  dashed lines show the two-body break-up.}
\end{figure}
\begin{figure}[ht]
\begin{center}
    \epsfig{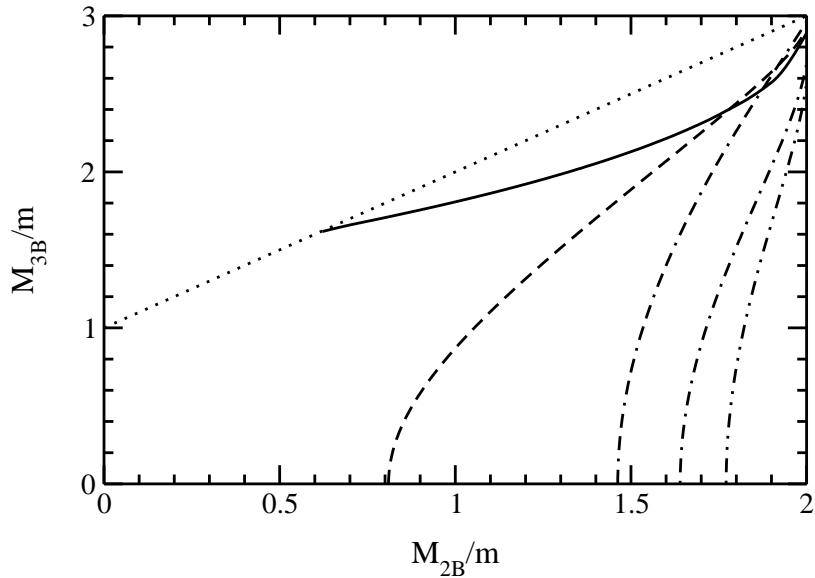}
\end{center}
\caption{\label{fig:M23}
  Three-body bound state as a function of $M_{2B}$ for different
  regularization schemes. Scheme A solid line. Others use invariant
  cut-off with different $\Lambda$: $\Lambda=4m$ (dash),
  $\Lambda=6m$ (dash-dot) $\Lambda=8m$ (dash-dot-dot)
  $\Lambda\rightarrow\infty$ (dash-dash-dot).}
\end{figure}
\begin{figure}[ht]
\begin{center}
    \epsfig{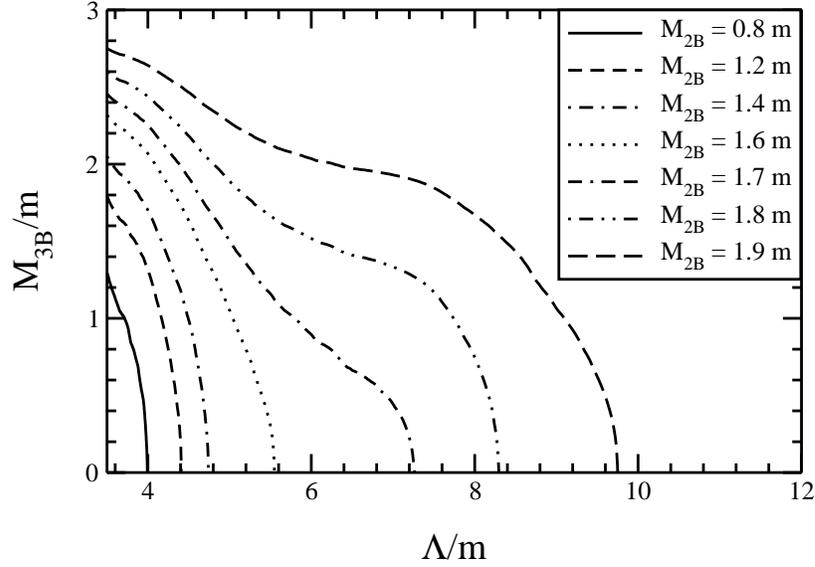}
\end{center}
\caption{\label{fig:B3L}
  Three-body bound state as a function of regularization masses
  $\Lambda$ for different $M_{2B}$. Binding energies
  $M_{2B}=0.8,1.2,1.4,1.6,1.7,1.8,1.9m$ (lines left to right)}
\end{figure}
\begin{figure}[ht]
\begin{center}
    \epsfig{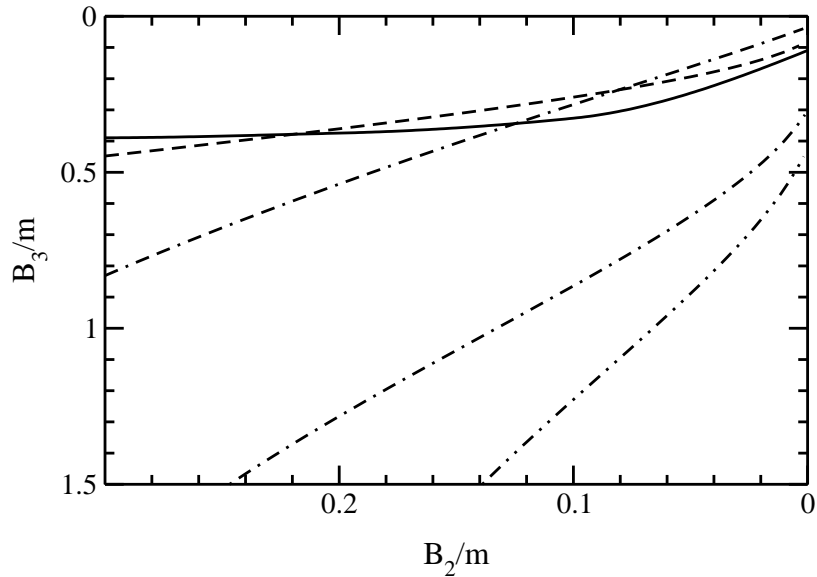}
\end{center}
\caption{\label{fig:B23}
  Binding energies $B_3(B_2)$. Line coding as in  Fig.~\ref{fig:M23}.}
\end{figure}

\subsection{General Discussion}
For the more general discussion we present our results as a function
of the interaction strength $\lambda$.  To solve (\ref{eqn:fad}) we
use the cut-off parameters $\Lambda=4m,6m,8m,$ where $m$ is the
constituent mass, and $\Lambda/m=10^{15}\rightarrow\infty$.  In
Fig.~\ref{fig:m23L} we show $M_{2B}$ and $M_{3B}$ (in units of m) of
the solution of the two-body bound state and the three-body bound
state as a function of the strength $\lambda$. Several features can be
recognized:

{\em i)} The values of the strength parameter $\gl,$ where bound
states exist, become smaller as the cut-off $\gL$ becomes larger. This
can been seen clearly, if one assumes a two-body bound state $M_{2B}$
to exist, see (\ref{eqn:Bound}): As $\gL\rightarrow \infty$ and
therefore $B_2(M_{2B})\rightarrow \infty$ the strength $\gl\rightarrow
0$, provided the mass of the bound state is independent of the
cut-off.  In turn, this implies that the interaction strength
depends on the cut-off, $\lambda\rightarrow\lambda_\Lambda$.

{\em ii)} For a certain $\lambda_\Lambda$ the three-body bound
state shows $M_{3B}\rightarrow 0$ despite a finite value of $M_{2B}$.
This is the relativistic analog of the Thomas
collapse~\cite{Thomas1935}.

{\em iii)} As $M_{2B}\rightarrow 2m$, i.e.  reaches the continuum
state, the three-body bound state still exists.  The parameter for
which this situation occurs depends on the range of the interaction.
It may give rise to the Efimov effect, an issue not further addressed
here.

{\em iv)} There is a region of parameters where both $M_{2B}$ and
$M_{3B}$ exist. For this particular case it is possible to plot
$M_{3B}$ vs. $M_{2B}$. This has been done in previous calculations.

In fact, for case {\em iv)} we show the respective plot
$M_{3B}(M_{2B})$ in Fig.~\ref{fig:M23}. As explained before, using
(\ref{eqn:Bound}) as a physics input, it is possible to eliminate the
$\lambda_\Lambda$ and $\Lambda$ dependence from the solution of the
three-body equation. 

A similar plot that is given in Fig.~\ref{fig:B3L} shows the cut-off
dependence of $M_{3B}(\Lambda)$ for a given fixed two-body mass
$M_{2B}$.  Again the mass $M_{3B}\rightarrow 0$ indicates the Thomas
collapse.

The region close to threshold is given in
Fig.~\ref{fig:B23} and parameterized in terms of the binding energies
\begin{eqnarray}
  B_3&=&m+M_{2B}-M_{3B}\\
  B_2&=&2m-M_{2B}.
\end{eqnarray}

It is worthwhile to note that for large invariant cut-off masses the
resulting function $M_{3B}(M_{2B})$ coincides with the result that
utilizes (\ref{eqn:Bound}) and solves (\ref{eqn:fad}) without cut-off
restrictions. In this sense we reproduce the result of
Ref.~\cite{Carbonell:2002qs}. Here we go further, extending our
regularization procedure and solving the three-body equation without
assuming a two-body bound state. To this end the introduction of a
cut-off is adequate. On the other hand, a cut-off $\Lambda$ can be
considered a physical input, e.g. as a scale limit for physics beyond
the ``zero range approximation'' or as a range parameter of the
interaction.  

In Fig.~\ref{fig:M23} we have also given the earlier result using the
regularization scheme A explained in the previous section. This shows
no Thomas collapse, because of the additional restriction $M_2^2\ge 0$
for the intermediate state~\cite{Frederico:1992uw}.

\subsection{Introducing a scale}

\begin{figure}[t]
\begin{center}
    \epsfig{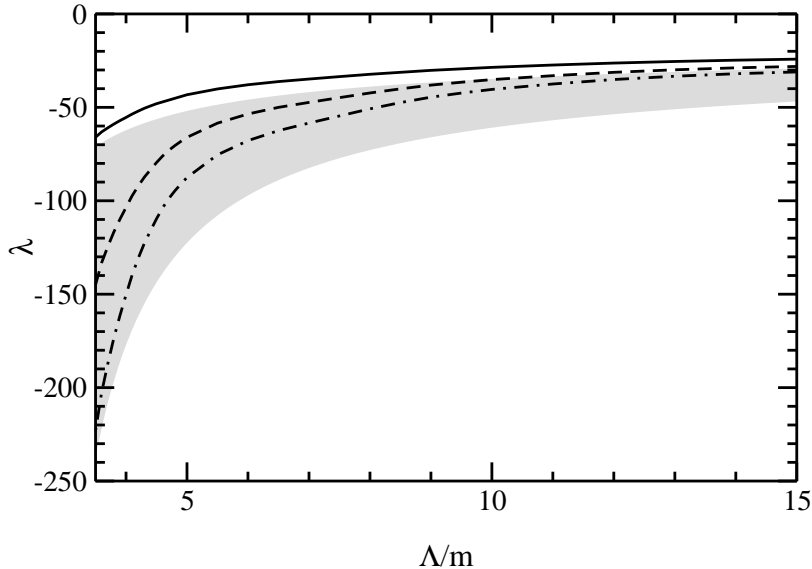}
\end{center}
\caption{\label{fig:L} 
  $\lambda(\Lambda)$ from a fit of $M_{3B}$ to the proton mass, with
  $m=315$ MeV (solid line), $m=400$ MeV (dashed line) $m=900$
  MeV (dash-dot). Grey area bounded by $M_2=2m$ (upper), $M_2=0$
  (lower).}
\end{figure}
\begin{figure}[t]
\begin{center}
    \epsfig{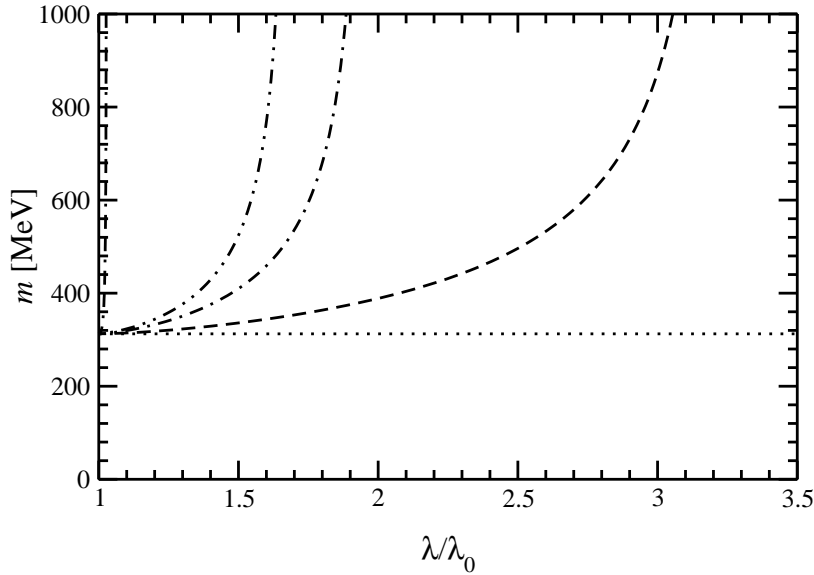}
\end{center}
\caption{\label{fig:qmass}
  Quark masses from a fit of $M_{3B}$ to the proton mass as a function
  of the strength $\lambda/\lambda_0$ for different cut-off $\Lambda$,
  $\lambda/\lambda_0=1$ as explained in the text. The dotted
  horizontal line indicates $m_p/3$. Other lines as in Fig.~\ref{fig:M23} }
\end{figure}
In order to achieve more quantitative results we now introduce a scale
into the calculation. Although the model is not elaborate enough to
expect a complete description of the baryon dynamics we choose the
proton mass $m_p=938$ MeV.  This seems natural for a three-quark
system and has been used before in this
context~\cite{deAraujo:1995mh}. For a more detailed description the
spin has to be included on the light front, e.g., along the lines of
Ref.~\cite{Beyer:1998xy}.  The scale restricts the parameter space
$(m,\lambda,\Lambda)$ resulting in a two-dimensional surface of
allowed parameters. To be more physical we present our results in two
ways. First we choose a specific quark mass (A) and second we choose a
specific cut-off $\Lambda$ (B).

A. As constituent masses we choose $m\simeq m_\rho/2\simeq
400$MeV~\cite{deAraujo:1995mh} and $m\simeq m_p/3\simeq 315$MeV and a
rather large one of $m=900$ MeV. By fitting the proton mass this
determines functions $\lambda(\Lambda)$ that are shown in
Fig.~\ref{fig:L}. The grey area indicates the parameter values for
which a two-body bound state exists.  There are three interesting
effects. {\em i)} For the weakly bound system there exists a
``critical'' range (parameterized through $\Lambda$) of the potential
where no two-body bound state exists (lines outside the grey area).
This is the case for $m=315$ MeV for all values of $\Lambda$ and for
$m=400$ MeV for values of $\Lambda$ above $\sim 9.6m$. Hence there is
a possibility that the nucleon could be a Borromean system.  For the
nonrelativistic case of Borromean systems, see
e.g.~\cite{Moszkowski:2000ni}. {\em ii)} An intermediate quark mass
already leads to the possibility of bound two-quark states.  However,
there is a moderate dependence on the cut-off, i.e. range of the
interaction, and possibly other details of the dynamics.  {\em iii)} A
large quark mass ($m=900$ MeV) leads to a large ``binding energy'' and
hence requires the existence of bound two-quark states.

B. Alternatively one may choose a particular cut-off $\Lambda$.  This way,
from $m_p=M_{3B}$ we find a relation $m(\lambda)$ that of course still
depends parametrically on $\Lambda$. In Fig.~\ref{fig:qmass} we show
the resulting quark mass dependence necessary to reproduce the proton
mass. To compare different cut-offs we have normalized the strength to
the threshold strength $\lambda_0$, i.e. were ($M_{3B}=3m$). The dotted
horizontal line indicates the continuum limit $m=m_p/3$. 

Finally, one may wish to also fix the last parameter,
$\Lambda$. To this end one could use more empirical information. This
has been done in \cite{deAraujo:1995mh} for the regularization scheme
A. We do not pursue this further. A detailed study of the nucleon
dynamics is beyond our present aim and the capability of the model in
its present form.

\section{Conclusion}
We have investigated in detail the stability of a relativistic three
boson system subject to an effective zero range interaction. We find
the relativistic analog of the Thomas collapse, which is given by
$M_{3B}\rightarrow 0$. Hence, the Thomas collapse is present also for
the relativistiv case, but depends on the regularization scheme. This
could be seen by introducing an invariant cut-off $\Lambda$ that
restricts the virtual masses in the respective few-body equations. The
invariant cut-off can be understood as the limiting scale for physics
beyond ``zero-range approximation'' or as an effective size parameter.
Therefore a finite value for $\Lambda$ is meaningful. In the limit
$\Lambda\rightarrow \infty$ we reproduce results given
earlier~\cite{Carbonell:2002qs}.  For a more quantitative discussion
we have introduced the nucleon mass as a scale~\cite{deAraujo:1995mh}.
Further analysis including a  treatment of
spins~\cite{Beyer:1998xy} is left for future investigations.

Acknowledgement: Work supported by Deutsche Forschungsgemeinschaft.

\end{document}